\documentclass[%
reprint,
superscriptaddress,
preprintnumbers,
nofootinbib,
amsmath,amssymb,
aps,
prd,
]{revtex4-1}

\usepackage{hyperref}
\hypersetup{colorlinks=true, citecolor=blue, urlcolor=blue, linkcolor=blue}
\usepackage[T1]{fontenc}
\usepackage{float}      		% figure positioning
\usepackage{dcolumn}    		% Align table columns on decimal point
\usepackage{bm}         		% bold math
\usepackage{graphicx}		% Including figure files
\usepackage{amsmath}		% Advanced maths commands
\usepackage{amssymb}		% Extra maths symbols
\usepackage{aas_macros}
\usepackage{xcolor}
\usepackage{newtxtext,newtxmath}
\usepackage[linesnumbered,ruled,vlined]{algorithm2e}
\usepackage{algpseudocode}
\usepackage{tikz}
\usetikzlibrary{fit,calc}

\begin{document}
\title{Improving Convolutional Neural Networks  for Cosmological Fields with Random Permutation}

\author{Kunhao Zhong}\email{kunhaoz@sas.upenn.edu}
\affiliation{Department of Physics and Astronomy,
University of Pennsylvania, Philadelphia, PA 19104, USA}

\author{Marco Gatti}
\affiliation{Department of Physics and Astronomy,
University of Pennsylvania, Philadelphia, PA 19104, USA}

\author{Bhuvnesh Jain}
\affiliation{Department of Physics and Astronomy,
University of Pennsylvania, Philadelphia, PA 19104, USA}

\date{\today}

\begin{abstract}
Convolutional Neural Networks (CNNs) have recently been applied to cosmological fields -- weak lensing mass maps and galaxy maps. However, cosmological maps differ in several ways from the vast majority of images that CNNs have been tested on: they are stochastic, typically low signal-to-noise per pixel, and with correlations on all scales. Further, the cosmology goal is a regression problem aimed at inferring posteriors on parameters that must be unbiased. We explore simple CNN architectures and present a novel approach of regularization and data augmentation to improve its performance for lensing mass maps. We find robust improvement by using a mixture of pooling and shuffling of the pixels in the deep layers. The random permutation regularizes the network in the low signal-to-noise regime and effectively augments the existing data. We use simulation-based inference (SBI) to show that the model outperforms CNN designs in the literature. We find a 30\% improvement in the constraints of the $S_8$ parameter for simulated Stage-III surveys, including systematic uncertainties such as intrinsic alignments. We explore various statistical errors corresponding to next-generation surveys and find comparable improvements. We expect that our approach will have applications to other cosmological fields as well, such as galaxy maps or 21-cm maps.

\end{abstract}

\maketitle

\section{Introduction}

The Large Scale Structure (LSS) of the universe contains crucial information about its late-time growth history and fundamental physics. Over the past few decades, a diverse array of probes has been employed to study these structures, each contributing significantly to our understanding of the universe. Prominent among these are the Stage-III LSS surveys, which include the Dark Energy Survey (DES)~\cite{DES:2016jjg,DES:2017qwj,DES:2021bvc}, the Kilo-Degree Survey (KiDS)~\cite{Kuijken:2015vca,Heymans:2020gsg,KiDS:2020suj}, the Hyper Suprime-Cam Subaru Strategic Program (HSC)~\cite{2023arXiv230400701D, 2023arXiv230400702L, 2023arXiv230400705S}, and the Baryon Oscillation Spectroscopic Survey (Boss and eBOSS)~\cite{Smee2013,2022MNRAS.511.5492Z,eBOSS:2020yzd}. 

The upcoming Stage-IV LSS surveys, including Rubin Observatory's LSST~\cite{LSST:2008ijt}, the Roman Space Telescope~\cite{Dore:2019pld}, and the recently launched Euclid mission~\cite{laureijs2011euclid}, in combination with the Dark Energy Spectroscopic Instrument~\cite{2019BAAS...51g..57L}, are or will be gathering data soon. All these efforts will shed light on two of the most mysterious problems in physics: dark matter and dark energy.

One of the most important analytical tools used in these studies is the two-point correlation function. This function links observational measurements with theoretical predictions derived from perturbation theory. However, the two-point correlation function only captures the Gaussian information. 

A variety of non-Gaussian statistics have thus been proposed to study cosmological fields. These include the three-point function and bispectrum~\cite{10.1046/j.1365-8711.2003.06868.x, 2023JCAP...10..028H, 10.1111/j.1365-2966.2004.07410.x}, peaks and voids~\cite{2000ApJ...530L...1J, 2009ApJ...698L..33M, 2018MNRAS.474..712M, 2021MNRAS.506.1623H, 2022MNRAS.511.2075Z, 2023arXiv230810866M}, Minkowski functionals~\cite{2012MNRAS.419..536M, 2012PhRvD..85j3513K, 2022OJAp....5E..13G}, Betti numbers and persistent homology~\cite{2019JCAP...09..052F, 2021A&A...645A.123P, 2021A&A...648A..74H, 2022A&A...667A.125H}, k-NN and CDF~\cite{2023MNRAS.519.4856B, 2023MNRAS.526.5530A}, and wavelet transform based statistics~\cite{2020MNRAS.499.5902C, 2021arXiv211201288C, 2022PhRvD.106j3509V, 2023arXiv231100036H, 2022mla..confE..40P, 2023arXiv231017557G}. See Ref.~\cite{2023A&A...675A.120E} for a forecast with the settings of the Euclid Mission.

Most recently, machine learning has been used to extract cosmological information at the field level, particularly with Convolutional Neural Networks (CNN)~\cite{Fluri2018, Ribli2019, 2018PhRvD..97j3515G, 2019PhRvD.100f3514F, 2019NatAs...3...93R, 2020PhRvD.102l3506M, 2021MNRAS.501..954J, 2022MNRAS.511.1518L,  2023MNRAS.521.2050L, 2020PhRvD.102j3509H, Akhmetzhanova:2023hiy}, Graph Neural Networks (GNN)~\cite{2019A&C....27..130P, 2022PhRvD.105h3518F, 2022OJAp....5E..18M, 2022MLS&T...3aLT03M}, and generative models~\cite{Dai:2022dso, 2021A&A...651A..46U,Yiu:2021pga, CAMELS:2020cof, Dai:2023lcb}.

Neural networks generally surpass traditional Gaussian statistics in constraining cosmologies, such as the power spectrum. However, the extent of the improvement can vary significantly depending on the realism of the simulations and the scales considered. When compared to other non-Gaussian statistics, the situation is more complicated. For instance, several studies~\cite{Fluri2018, 2018PhRvD..97j3515G} have suggested that CNNs outperform peak counts methods, whereas others have pointed out that their enhancement over Minkowski functionals is only moderate~\cite{2020PhRvD.102l3506M}. Moreover, there are scenarios where CNNs appear not to perform as effectively as other methods, such as the scattering wavelet transform~\cite{2020MNRAS.499.5902C}. This observation raises the question that while neural networks offer substantial advancements in many areas, their efficacy can vary for cosmological fields. In other words, even though the CNN often contains thousands of free parameters to approximate the underlying function, it might not be suitable to directly apply the commonly used classification-oriented CNN in cosmology. Such insights are crucial for guiding future research and methodology choices in the field of cosmology, particularly in the integration and application of machine learning techniques.

In this paper, we explore modifications to the architecture of CNNs to enhance their performance in cosmological tasks. Cosmological maps such as weak lensing mass maps differ from images used to develop CNNs in that they are stochastic and low signal-to-noise. Moreover  cosmological inference we use them for regression, not classification as is typically done in industry. So we explore a variety of regularization and data augmentation schemes in applying CNNs on lensing maps. Our preliminary findings suggest that employing a combination of maximum and average pooling  improves the network's ability to extract relevant information. We also introduce a novel regularization scheme rooted in our understanding of cosmological fields: the incorporation of random permutation layers designed to effectively augment the training data by sacrificing the information of large-scale correlation. We demonstrate the effectiveness of this straightforward technique for statistical noise levels expected in current and upcoming weak lensing surveys. We also include a number of sources of systematic uncertainty. 

Additionally, we conduct comparative analyses with conventional regularization methods and alternative strategies that disrupt large-scale correlation. Our initial results indicate that the permutation operation directly applied to the latent space of the neural network consistently boosts accuracy in both the simplified and realistic simulation cases considered in this work. This suggests that by selectively filtering out large-scale information in the deeper layers, the neural network exhibits enhanced performance in extracting information from more critical scales. We anticipate that this simple regularization technique not only holds promise for improving weak lensing fields but may also for other stochastic fields, such as galaxy maps or 21-cm maps. We do not assert that our architecture represents the optimal model as we have not carried out a systematic study for all possible use cases.

The structure of this paper is as follows: In Sec.~\ref{sec:methodology} we describe the simulations we used for this work, the preprocessing of the data, and the simulation-based inference setup. In Sec.~\ref{sec:CNN_for_cosmology} we discuss how the CNN for cosmological fields should be different from the standard design for image classification. In Sec.~\ref{sec:CNN_with_shuffle} we present the random permutation layer as a regularization scheme that helps model generalization. We summarize the findings and outlook for future work in Sec.~\ref{sec:Conclusion} and we perform extensive tests comparing to other models in the Appendices.

\section{Methodology}
\label{sec:methodology}

\subsection{Simulation of Weak Lensing Convergence Field}

\begin{figure}
\centering
\includegraphics[width=\columnwidth]{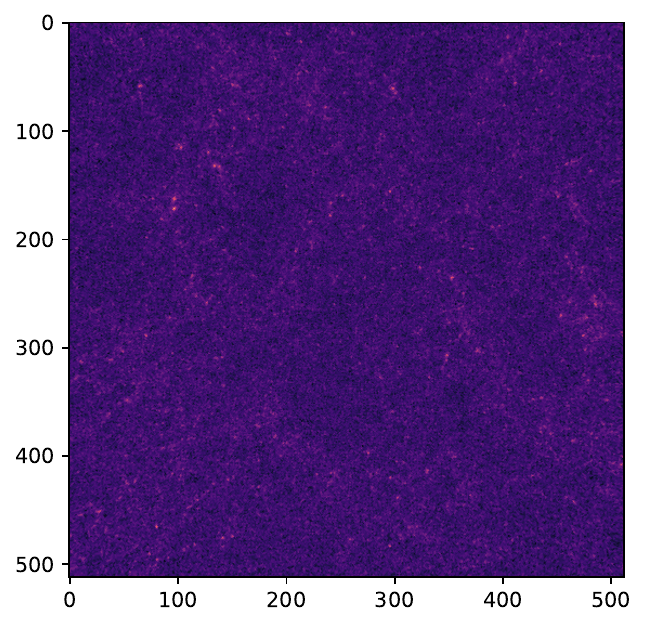}
\includegraphics[width=\columnwidth]{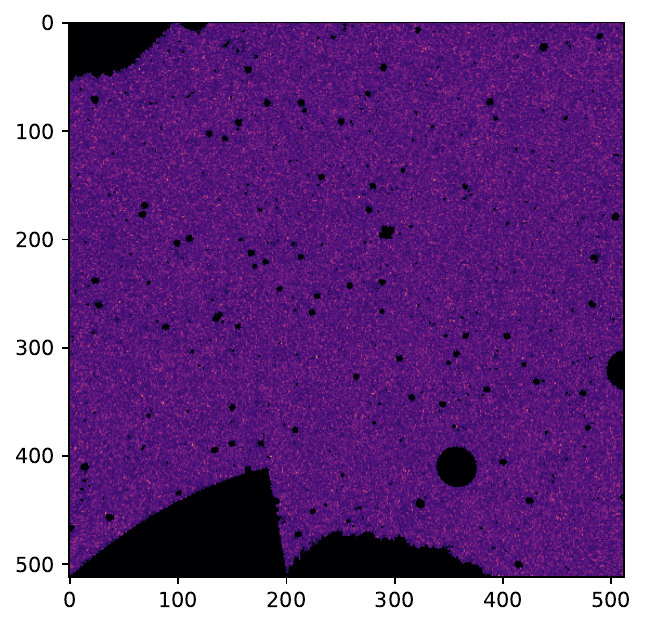}
\caption{
\textit{Upper Panel}: Example patch of the weak lensing mass maps for simulation case 1, in which no mask is applied and only one tomographic bin is used. The side length is 58 degrees and is given in pixel units. \textit{Lower Panel}: Example patch for simulation case 2, which includes a mask that follows the DES footprint. The figure shows the third of the four tomographic bins.
} 
\label{fig:example_patch} 
\end{figure}

For this work, we use weak lensing mass maps (i.e., maps of the weak lensing convergence field) created from the \texttt{DarkGridV1} N-body simulation suite \citep{Zuercher2021,Zuercher2021b}. The \textsc{DarkGridV1} suite consists of \( \Lambda \)CDM-only simulations that vary two parameters, \( \Omega_{\rm m} \) and \( \sigma_8 \), exploring 58 different cosmologies~\footnote{The samples are chosen to follow lines of approximately constant $S_8$. See Fig. 2 of Ref.~\cite{Zuercher2021} for the grid spanning the $ \Omega_\mathrm{m} - \sigma_8 $ plane.}. Each cosmology is represented by five independent full-sky simulations. The simulations have been produced using the  \textsc{PKDGRAV3} code \citep{potter2017pkdgrav3}; the code produces particle number counts at different redshifts ($100$ redshifts from $z=49$ to $z=0.0$), provided as \textsc{HEALPIX} \citep{GORSKI2005} maps. For this work, we downsample the resolution of the original maps to \textsc{NSIDE = 512}, corresponding to a pixel resolution $\approx$ 6.9 arcmin. To create weak lensing mass maps, we follow a procedure similar to that in \cite{sourceclustering,Gatti2023}. First, noiseless convergence maps are produced from the particle counts for each redshift shell assuming the Born approximation \citep{Fosalba2015}. Noiseless shear maps, for each redshift shell, are obtained from the convergence maps using a generalisation of the Kaiser-Squires algorithm \citep{y3-massmapping, 1993ApJ...404..441K}. Then, we consider two cases:

\textbf{simulation case 1 - no observational systematics or tomography}. In case 1, we simplify the map-making procedure, and we choose to not include observational systematics in the modeling. We first generate integrated DES-Y3-like shear maps, by weighting the shear maps as a function of redshift by the DES-Y3 redshift distributions \cite{y3-sompz}. To simplify and reduce the number of maps generated, we focus on just one of the four DES redshift bins, specifically the third bin. Next, we introduce Gaussian noise to the shear maps. This noise is assumed to have a per-pixel value of $\sigma = \sigma_e / \sqrt{N_{\text{eff}}}$, where $\sigma_e = 0.26$ represents the shape noise, and $N_{\text{eff}} = 5.6 \times A_{\text{pixel}}$ is the effective number of galaxies per pixel. This is equivalent to adding $\sigma = \sigma_e / \sqrt{2 N_{\text{eff}}}$ to the mass map as used in Ref.~\cite{Ribli2019}. Note that here we have used the number density 5.6 of the full DES-Y3 shear sample, not just that of the third bin. This approach is intended to yield cosmological constraints similar to the entire DES-Y3 survey, even though we are using only one tomographic redshift bin. After adding the noise to the full-sky convergence map, we cut 12 non-overlapping square patches; each patch is $512\times 512$  pixels (see Fig. \ref{fig:example_patch}), and is chosen to be centered on the center of a \textsc{HEALPIX} pixel with a resolution \textsc{NSIDE = 1}. For the 285 full-sky simulations we generated, we have $285\times12=3420$ patches in total. 

\textbf{simulation case 2 - full map-making procedure}. In case 2, we produce weak lensing maps using the full procedure implemented in \cite{sourceclustering,Gatti2023}. The procedure produces four maps, one for each of the four DES tomographic bins. Moreover, a DES footprint (fixed mask for bright galaxies, see Fig.~\ref{fig:example_patch}) is applied. Each map also includes observational and astrophysical systematics, namely redshift uncertainties, shear biases, intrinsic alignment (in the form of the non-linear alignment (NLA) model), and source clustering. These are managed using various nuisance parameters. Practically, for each map, we forward-model these effects by randomly selecting the nuisance parameters that control them from their respective priors -- see \cite{Gatti2023} for details. In this case, we only train using a single patch of the same galaxy mask, approximately centered at the center of the DES footprint (see Fig. \ref{fig:example_patch}). For each full sky simulation with the same cosmology, we shift and rotate the DES footprint four times to generate 4 non-overlapping maps. For the 2280 full-sky simulations we generated, we thus have $2280 \times 4 = 9120$ patches in total.

In both cases, the training set is images of size 512x512 with a resolution of 6.9 arcmin, which corresponds to $3367 \mathrm{degree}^2$. In case 1 images have one channel, while in case 2 images have four channels, one for each tomographic bin. These maps are significantly larger than previous works~\cite{Fluri2018, Ribli2019, 2018PhRvD..97j3515G, 2020MNRAS.492.5023J, 2022MNRAS.511.1518L, 2020PhRvD.102l3506M, 2023MNRAS.521.2050L} allowing us to use one patch to represent the survey area. Using a larger patch also has the benefit of keeping more modes. We tested the 6-layer CNN model in comparison to cutting to smaller patches of size 256x256, or 128x128. Although the total training set is larger with smaller patches, we observed a lower accuracy of the CNN, especially in the case of galaxy masks. Therefore, we use the large 512x512 image in this work. Note that the large sky area of a single projection breaks the flat-sky approximation. This in theory would not produce a biased result since it is part of the data compression for simulation-based inference as discussed in Sec.~\ref{sec:sbi}. However, in the future we will extend the work to the curved sky, e.g. with structures as presented in \cite{2019A&C....27..130P}.

\subsection{Data Augmentation and Training}\label{sec:data_augmentation_and_training}

Prior to training, we add random Gaussian noise to each pixel as a data augmentation technique, particularly effective for noisy images~\cite{Shorten2019}. In our methodology, we apply Gaussian noise with a mean equal to the image's mean value and a standard deviation equivalent to 10\% of the original image's standard deviation. In this work, we add this to all images (training, validation, and test). The noise level remains almost unchanged since it increases by $\sqrt{1+0.1^2}$. Note that since the additional noise is added according to each image and each channel, it is different from the global shape noise where the variance is fixed for all maps. While Ref.~\cite{Ribli2019} demonstrates the efficacy of augmenting data through random rotation and flipping, we observe no significant difference with these methods. The difference may be attributed to our approach of extracting patches from non-overlapping regions of the full sky. Although our data augmentation strategy differs from previous studies, we emphasize its importance as it improves the test error by $\sim 25\%$. This approach is also applied to the entire masked map, ensuring that the augmented map assigns non-zero values to the masked regions. This procedure can be optimized for future exercise. For example, using a dynamic approach such that different noise is added while training, the strengths of the noise can also decay away as the model is trained for more epochs.

We train separate CNNs for inferring $S_8$ and $\Omega_\mathrm{m}$, as the training time is short (15 mins for simulation case 1, and 1 hour for simulation case 2 on one A100 GPU). This approach also mitigates the issue of degeneracy when inferring two parameters simultaneously in a single network. In this work, we chose Mean Square Error (MSE) as the loss function, and we report both the Root Mean Square Error (RMSE) and the coefficient of determination (\textit{R}-squared) in the test results, which provides complimentary information of goodness of fit. Although MSE is widely adopted, alternative loss functions can significantly influence results. For instance, had we noticed a tendency for biased predictions at extreme input values, adopting a $\log$ function could have been beneficial to more heavily penalize such discrepancies. Since we did not have such biases for the prediction, we kept MSE in this work. The data set is randomly divided into 80\% for training, 10\% for validation, and 10\% for testing. The last 10\% test set data is then used for the simulation-based inference.

We note that in~\cite{Ribli2019} a more careful division method based on the initial conditions of simulations is utilized. Since our simulation only has 58 different cosmologies, we expect this not to be an issue as the validation set effectively covers the full sampling regions. Due to computational resource limits, we did not show every test with averaging over different random seeds. The random seeds in training affect the weight initialization and also the training-validation split, and thus affect the final results. However, we tested in a few cases with different random seeds and found the variance of RMSE to be as small as $\Delta \rm RMSE \sim 0.01\%$, which is partially because we trained for a long enough of epochs.

The CNN is implemented using \textsc{Pytorch}~\cite{pytorch} and trained using the \textsc{AdamW}~\cite{AdamW} optimizer. We use a batch size of 16 and a total epoch of 200. A learning rate scheduler, \textsc{ReduceLROnPlateau}, is used with a reduction factor of 0.3 (default is 0.1). For all other hyperparameters not explicitly mentioned, we use \textsc{Pytorch-v1.13}'s default settings. We hold hyperparameters with respect to training the same for all tests in this paper. We observe effective convergence with this substantial number of epochs. However, it is noteworthy that Quasi-Newtonian optimization methods, such as L-BFGS~\cite{andrew2007scalable} or the Levenberg–Marquardt algorithm~\cite{LEVENBERG-KENNETH}, are theoretically more suitable for parameter inference where accuracy is more important. Yet, their application often demands substantial memory due to the Jacobian computation. An effective Quasi-Newtonian optimizer for precision training is an interesting direction for future research.

\subsection{Simulation-based Inference}\label{sec:sbi}

\begin{figure*}
\centering
\includegraphics[width=2\columnwidth]{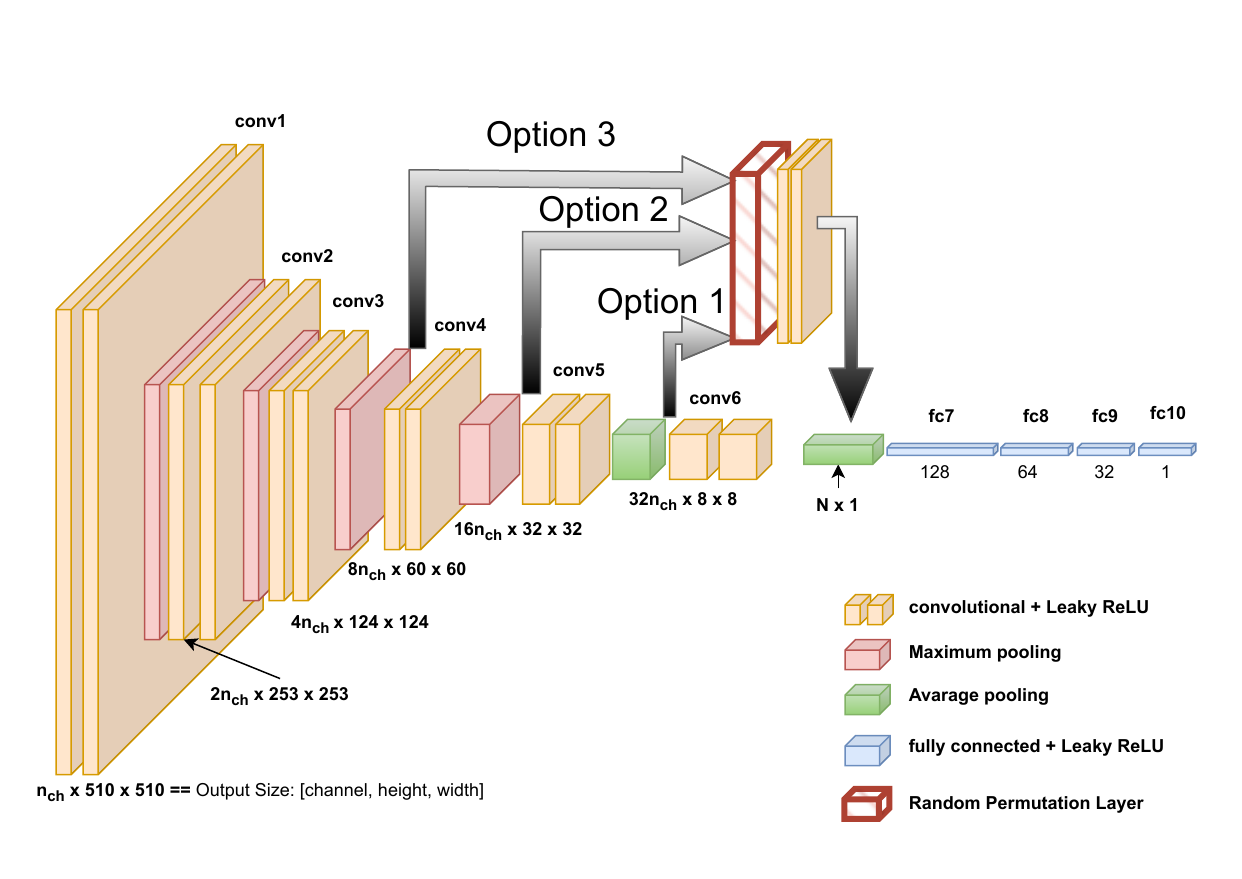}
\caption{The 6-layer CNN design used in our study, including the optional random permutation blocks.  The network is designed similarly to a VGG-net~\cite{2014arXiv1409.1556S}. In this work, we use $n_\mathrm{ch}=8$ channels in the first layer, with $n_\mathrm{ch}$ increasing by a factor of 2 for successive deeper layers. The three ''Options'' shown involve replacing the deeper layers with a random permutation layer followed by convolution and activation at different depths as indicated (note that the arrows do not indicate skip connections). In Option 1, one extra random permutation layer is added without any change to the 6-layer CNN design. For Option 2 and Option 3, the random permutation is performed at earlier layers.
Note that we also mix average and maximum pooling layers: see Sec.~\ref{sec:CNN_with_shuffle} for details. The blocks shown in this figure are not to scale.}
\label{fig:nn_design} 
\end{figure*}

We do not use Neural Networks to directly infer posteriors, for that one usually needs to assume a Gaussian likelihood. Instead, we use the Neural Density Estimator (NDE) package \textsc{PyDelfi} introduced in~\cite{2018PhRvD..97h3004C}. Our CNN thus serves as a field-level data compressor similar to the \textsc{DeepCompressor} in~\cite{2021MNRAS.501..954J}. \textsc{PyDelfi} provides different flow models for Neural Density Estimation. In this paper, we use Masked Autoregressive Flows (MAF)~\cite{2017arXiv170507057P}, which is a stacked version of Masked Autoencoders for Distribution Estimation (MADEs)~\cite{2015arXiv150203509G} with normalizing flow. The network of the NDE is trained to minimize the Kullback-Leibler (KL) divergence~\cite{kullback1951information} between the true probability distribution and the proposed probability distribution. The true likelihood is of course unknown, but it is reduced to a constant in calculating the expectation value of the KL divergence over the implicit prior (the prior set by the distribution of the point cloud of parameter values used for the training simulations). For details, see~\cite{2018PhRvD..97h3004C, 2021MNRAS.501..954J, 2020PNAS..11730055C}.

In this work, we stack 2 MAFs with 2 and 3 hidden layers of length 50. Since one patch corresponds to approximately 3367 square degrees of the sky, we infer the posterior using only one image. Note that the simulations used for NDE training are the \textit{test set} that is not used for CNN training or validation. After the iso-likelihood surface is learned, we use the affine MCMC sampler \textsc{emcee}~\cite{2013PASP..125..306F} to get the final posterior with a flat prior of $S_8 \in [0.45, 1.1]$ and $\Omega_\mathrm{m} \in [0.1,0.5]$. For the purpose of forecasting, we average over 9 predicted values of the same underlying cosmology as the target vector (in our case $S_8$ and $\Omega_\mathrm{m}$). This way the posterior is more centered and minimizes the effect of the prior, which allows us to make a more direct comparison of the constraining power.

We follow previous work on the validation for SBI via the empirical coverage test~\cite{2017arXiv170604599G, 2021arXiv211006581H, 2023PMLR..20219256L}, which plots the expected coverage probability (usually from the highest posterior density) vs the credible region. For a calibrated posterior with enough samples of the coverage test, one should find that the true parameter is contained within the $X\%$ credible region for $X\%$ times of the tests. This plot, often called the \textit{p-p} plot, is not unique to SBI as it is used for comparing two probability distributions in general. In this work, we use \textsc{TARP} (Test of Accuracy with Random Points)~\cite{2023PMLR..20219256L}, which is a more computationally efficient way of performing the coverage test. It is important to note that the coverage test aims to verify that the posterior learned from NDE is neither overconstrained nor underconstrained. A failed result would suggest that the NDE network is not suitable for the given problem, e.g. leading to overfitting. As shown in Sec.~\ref{sec:CNN_with_shuffle}, the results we obtain are well calibrated.

SBI, also called likelihood-free inference, is an emerging field and has been applied to cosmology in several recent studies~\cite{2020PNAS..11730055C, 2021arXiv210104653L, Hahn:2023udg, Lemos:2023myd}. The results for the inferred likelihood have been tested for summary statistics that have an explicit form of the likelihood such as the power spectrum~\cite{2021MNRAS.501..954J, 2023arXiv231017557G, 2023MLS&T...4aLT01L}. However, a robust convergence criterion and calibration of misspecification are missing. Various other robustness tests have been proposed to check that the results are not biased or over-estimated~\cite{2022arXiv221109602L, 2024arXiv240102413J}. Since the main topic of this paper is to improve the CNN for data compression, we defer a detailed study of SBI in the presence of model misspecification to future work.

\section{CNN for Cosmological Fields}
\label{sec:CNN_for_cosmology}
\begin{algorithm*}[t]\label{algorith:1}
\begin{algorithmic}
\caption{Pytorch pseudo-code of CNN with random permutation}
\State \textbf{class} ShuffledCNN:
\State \quad \textbf{def} \_\_init\_\_(self):
\State \quad \quad \tcp{Initialize CNN layers here} 
\State

\State \quad \textbf{def} randomize\_images(self, tensor, image\_size):
\State \quad \quad x\_flat = tensor.view(-1, image\_size*image\_size)
\State \quad \quad idx = torch.stack([torch.randperm(image\_size*image\_size) for \_ in range(x\_flat.size(0))])
\State \quad \quad if tensor.is\_cuda:
\State \quad \quad \quad idx = idx.cuda()
\State \quad \quad x\_randomized\_flat = torch.gather(x\_flat, 1, idx)
\State \quad \quad x\_randomized\_flat.view(tensor.shape[0], tensor.shape[1], image\_size, image\_size)

\State
\State \quad \textbf{def} forward(self, x):
\State \quad \quad \tcp{Implement the initial forward pass here}
\State \quad \quad \tcp{ Randomly permute the feature maps at corresponding position}
\State \quad \quad if self.training: \tcp{This is optional}
\State \quad \quad \quad x = self.randomize\_images(x, x.shape[-1])
\State \quad \quad  x = self.conv6(x)
\State \quad \quad  x = self.LeakyReLU(x)
\State \quad \quad  x = self.pool6(x)
\State \quad \quad \tcp{Implement the rest forward pass here}
\State \quad \quad \textbf{return} x
\end{algorithmic}
\end{algorithm*}
Unlike the prevalent use of CNNs in image recognition tasks, their application in cosmology typically addresses regression problems rather than image classification. In this context, the precision of the output is most important. Moreover, the input data in cosmology, in particular weak lensing fields, significantly differs from conventional images, such as those of cats and boats and others that are part of ImageNet\cite{5206848}, the long-standing industry benchmark for image classification algorithms. As depicted in Fig.~\ref{fig:example_patch}, the weak lensing field is stochastic, low signal-to-noise, and has correlations on all scales. We seek a CNN approach that is simple and generalizable, rather than adopting complex architectures like ResNet~\cite{2015arXiv151203385H} or VGGNet~\cite{2014arXiv1409.1556S} that have been optimized for classification tasks on images in other domains. Prior research~\cite{Fluri2018, Ribli2019, 2018PhRvD..97j3515G, 2020MNRAS.492.5023J, 2022MNRAS.511.1518L, 2020PhRvD.102l3506M, 2023MNRAS.521.2050L} employing CNNs for analyzing weak lensing fields has yielded promising results using relatively simple architectures.

Our 6-layer CNN model, as shown in Fig.~\ref{fig:nn_design}, incorporates a notable departure from previous designs~\cite{Fluri2018, Ribli2019, 2018PhRvD..97j3515G, 2019PhRvD.100f3514F, 2019NatAs...3...93R, 2020PhRvD.102l3506M, 2021MNRAS.501..954J, 2022MNRAS.511.1518L,  2023MNRAS.521.2050L, 2020PhRvD.102j3509H}, where either the Maximum pooling or the Average pooling is used. Instead, we have combined the use of Maximum Pooling and Average Pooling layers. One motivation for us to use Maximum Pooling in the initial layers is driven by the observation that positive extreme values in our pixels correspond to galaxy clusters which are known to carry valuable cosmological information. Indeed, a study of CNNs using saliency maps for lensing mass maps ~\cite{2020PhRvD.102l3506M} found that clusters are more informative than voids in the presence of realistic noise. Conversely, deeper layers, correlating distant sky regions of the original map, may require equitable consideration. Here, our choice to switch to average pooling facilitates better generalization in the subsequent Multi-Layer Perceptron (MLP) layers. We test these choices as discussed below and in Appendix.~\ref{sec:test_pooling}.

We tested both simulation cases 1 and 2 and also increased the effective galaxy density (lower noise) to very approximately represent future LSST and Roman weak lensing surveys (we do not attempt to simulate the redshift coverage of those surveys). However, the best pooling option does not appear to correlate with noise level. In fact, the average pooling is similar to or better than the mix-pooling. When training on $\Omega_\mathrm{m}$ though, the network fails to converge. We also tested extreme pooling with the largest absolute value (which then includes underdense extrema) and found no difference in the final results. For this reason, we use the mixed-pooling as our baseline choice. One caveat is that the extremely valued pixels could also be the ones most affected by noise or limitations in the simulations or unmodelled effects; this becomes more of a concern for smaller pixel values than ours. Adopting pooling layers in a Bayesian way could potentially overcome this problem~\cite{10.1007/978-3-030-37548-5_10}.

It's important to note that this architecture is not optimal for every scenario,  such as when dealing with patches of different sizes or different choices of systematic uncertainties. Furthermore, we only optimize the architecture for the inference of $S_8$ while a better design for $\Omega_\mathrm{m}$ is likely different and merits follow-up work.

Nonetheless, our findings below show that simple, physics-inspired modifications to a CNN can significantly improve accuracy. This encourages the pursuit of tailored models for specific problems in cosmology, rather than relying on existing architectures primarily designed for image classification.

\section{CNN with Random Permutation Layers}\label{sec:CNN_with_shuffle}

\begin{figure}
\centering
\includegraphics[width=\columnwidth]{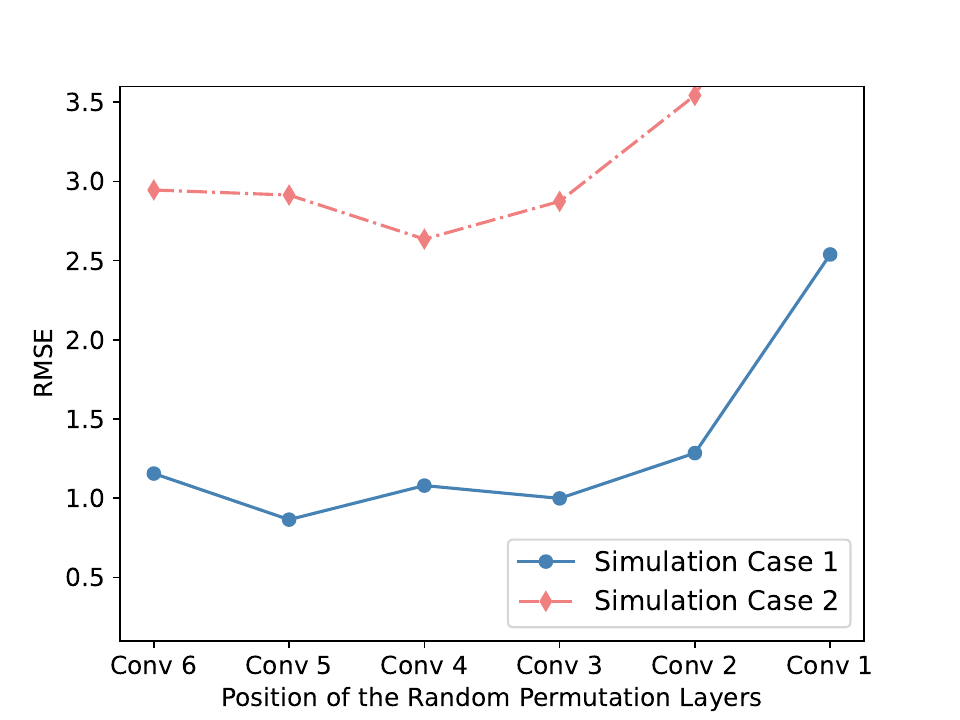}
\caption{The test errors for two simulation cases are shown for different positions of the random permutation layers. ''Conv 6'' denotes adding the shuffling operation before the final convolution, namely Option 1 in Fig.~\ref{fig:nn_design}. On the other hand, ''Conv 1'' refers to the first layer, so every pixel of the input image is randomly permuted (which leads to a higher loss as expected). While it is clear that shuffling is effective only in the deeper layers, the best shuffling position varies for the two cases and is not necessarily the final layers -- possibly because either option 2 or 3 leads to a simplified MLP structure (see  Fig.~\ref{fig:nn_design}). 
} 
\label{fig:cnn_loss_permute_position} 
\end{figure}
\begin{figure*}
\centering
\includegraphics[width=\columnwidth]{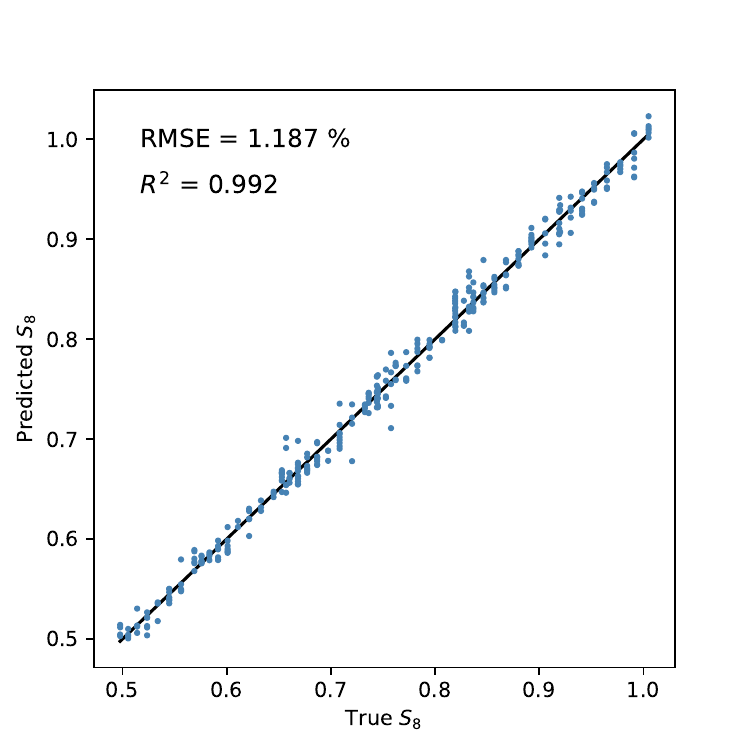}
\includegraphics[width=\columnwidth]{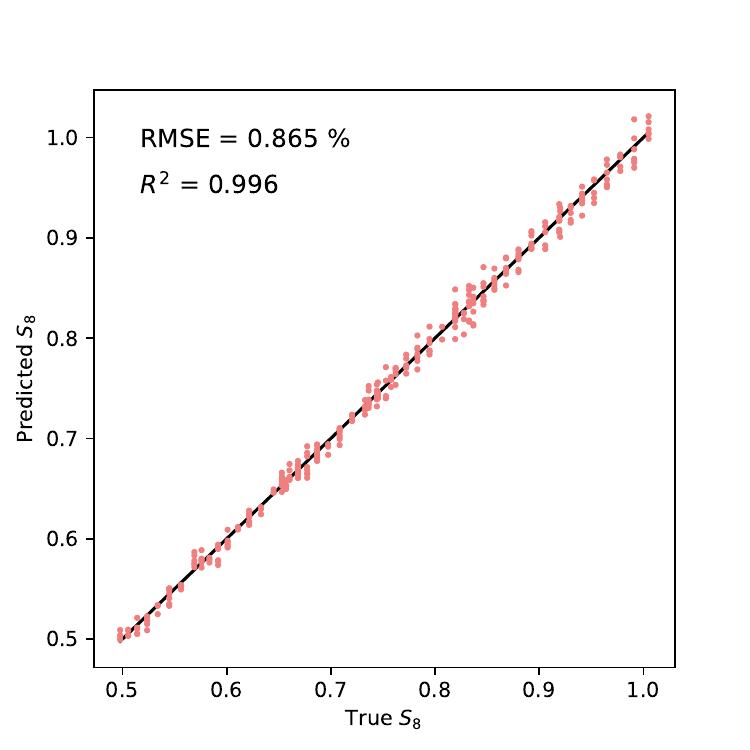}
\caption{Comparison of the test error in $S_8$ with and without shuffling (random permutation) for simulation case 1 (Case 2 is shown in Figure~\ref{fig:cnn_loss_case_2}). \textit{Left Panel}: 6-layer-CNN with no shuffle. \textit{Right Panel}: Shuffled-CNN (Option 2 in Fig.~\ref{fig:nn_design} which corresponds to a 5-layer CNN), which corresponds to removing correlations for scales larger than 109 arcmin. Note the significant improvement due to shuffling for the full range of values of $S_8$. 
The two panels in this figure represent the best results we found for non-shuffle and with shuffle.
} 
\label{fig:cnn_loss_case_1} 
\end{figure*}
\begin{figure*}
\centering
\includegraphics[width=\columnwidth]{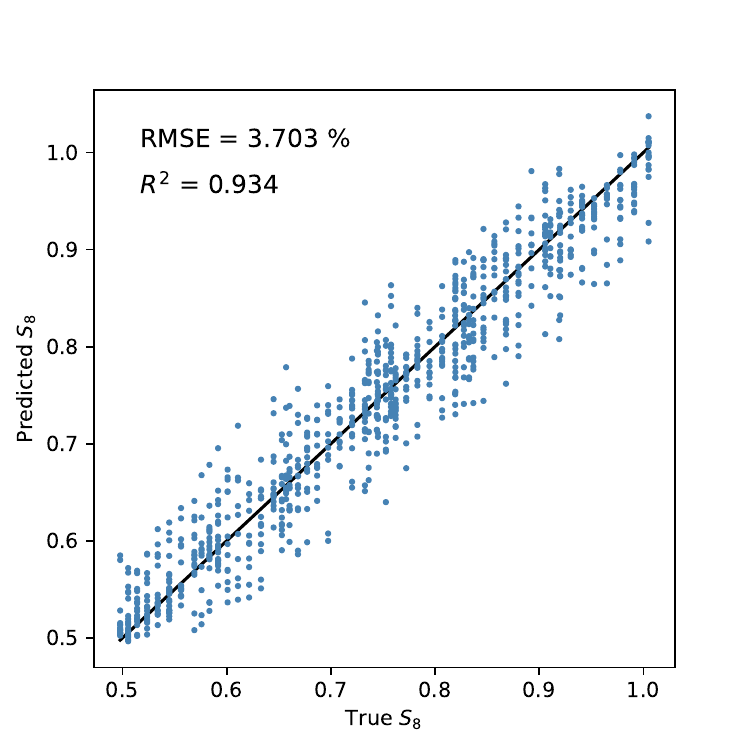}
\includegraphics[width=\columnwidth]{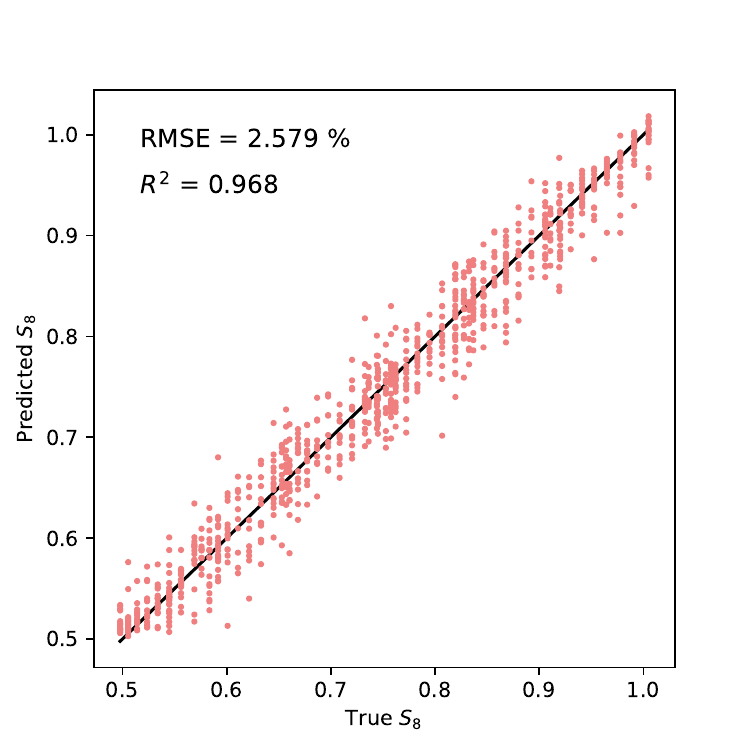}
\caption{Comparison of the test error in $S_8$ with and without shuffling for simulation case 2 (with mask and tomography). \textit{Left Panel}: 6-layer-CNN with no shuffle. \textit{Right Panel}: Shuffled-CNN (Option 3 in Fig.~\ref{fig:nn_design}which corresponds to a 4-layer CNN), which removes correlations for scales larger than 56 arcmin. The two panels in this figure represent the best results we found for non-shuffle and with shuffle. (If we shuffle at deeper layers, the performance is degraded to $\mathrm{RMSE} \approx 3.0\%$ but is still better than the non-shuffle case. )
}

%The direct comparison of the 6-layer CNN model should be option 1 since the only difference is the random permutation layer. Here we compare these two results because the accuracy for the models without shuffling decreases.} 
\label{fig:cnn_loss_case_2} 
\end{figure*}

Adding randomness is a common way in machine learning to improve the generalization accuracy. In this section, we introduce such a technique: random permutation of the pixels in the deeper layers of the network that correspond to large scales.

Figure~\ref{fig:nn_design} illustrates our design and the positions we replace the normal design with the random permutation layers. The random permutation layer shuffles (randomizes) the position of each pixel of the feature map. In this work, the shuffle is different for each batch, each channel, and changes every epoch. Note that the four tomographic bins have the same random permutation (shuffle) applied~\footnote{This is due to the fact that we use the Conv2d function implemented in \textsc{Pytorch}~\hyperlink{https://pytorch.org/docs/stable/generated/torch.nn.Conv2d.html}{https://pytorch.org/docs/stable/generated/torch.nn.Conv2d.html}}.

As an example, if we add the permutation layer before convolution layer 6 (Option 1), the input is the feature map of size 8 by 8 for 128 channels. The random permutation operation shuffles these 8x8 pixels randomly for each 128 channel differently. Each pixel in this feature map approximately corresponds to $512/8 \times 6.8 \approx 435.2\ \mathrm{arcmin}$. At this scale, the two-point correlation has a large sample variance. We find that by removing the information at large scales in the deeper layers, the neural network optimizes for the scales that are more relevant for cosmology. Moreover, it appears to extract some of the large-scale information in the earlier layers, though as discussed below we do not yet understand exactly how it does that. The shuffling operation can be inserted at different positions in the network, depending on the scale of which noise dominates over the signal. Note that this way of calculating scales is only approximate.  

We test the positions of the shuffling operation for simulation cases 1 and 2. As shown in Fig.~\ref{fig:cnn_loss_permute_position}, the accuracy depends on where we perform the shuffling. For simulation case 1 it is the second last layer and for simulation case 2 it is the third last layer. The simplification effectively makes them 5-layer and 4-layer models respectively. In general, we expect the results to depend on the underlying variation of cosmologies, the noise level, tomographic binning, patch size,  training set size, and other details. 
In particular, it will be interesting to see how increasing sample variance at large scales impacts the choice of the network~\cite{2022MNRAS.515.1942D}. 
To ensure a fair comparison, we also varied the depth of the non-shuffled CNN and found the 6-layer works the best (see Appendix.~\ref{sec:test_alternatives_to_random_shuffle}).

\begin{figure}
\centering
\includegraphics[width=\columnwidth]{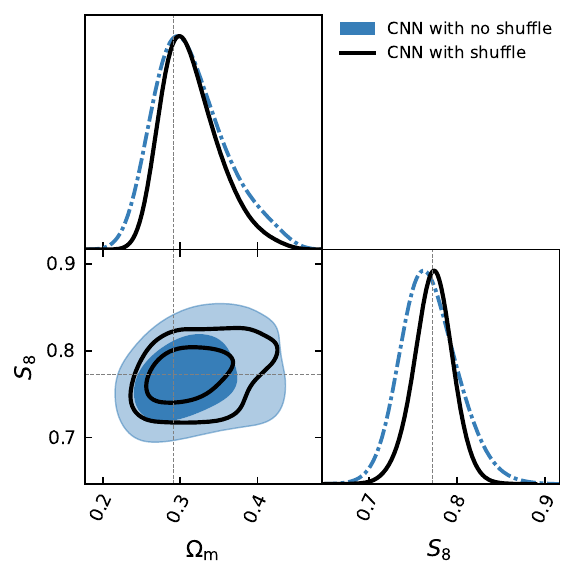}
\caption{The posterior inferred using simulation-based inference for simulation case 2. We find the improvement in the standard deviation of marginalized $S_8$ to be 31\%. The Figure of Merit $= 1 / \sqrt{\operatorname{det} \operatorname{Cov}\left(\Omega_{\mathrm{m}}, S_8\right)}$  is 80\% larger although this may not be the best metric as the posterior is non-Gaussian.} 
\label{fig:posterior} 
\end{figure}

The errors on the test set for both simulation cases are shown in Fig~\ref{fig:cnn_loss_case_1} and Fig~\ref{fig:cnn_loss_case_2} respectively. The posteriors for the two parameters are shown in Fig.~\ref{fig:posterior}. We find that the CNN with shuffling the standard deviation of $S_8$ is 31\% smaller than the non-shuffle case.

Note that the CNN for the two cases is optimized separately, and indeed it is also different for $S_8$ and $\Omega_\mathrm{m}$. In particular, it is the 6-layer model for non-shuffle for $S_8$, the 4-layer model for non-shuffle for $\Omega_\mathrm{m}$, and both 4-layer model for shuffle-CNN (option 3). To make the comparison for the same CNN model with the same weight size, we use shuffle option 1 for $S_8$ as well and get a 17\% improvement on the standard deviation of $S_8$. However, it is important to note that this change of layer depths only helps the shuffle-CNN, while the non-shuffle CNN suffers from insufficient depth in our case.

\begin{figure}[t]
\centering
\includegraphics[width=\columnwidth]{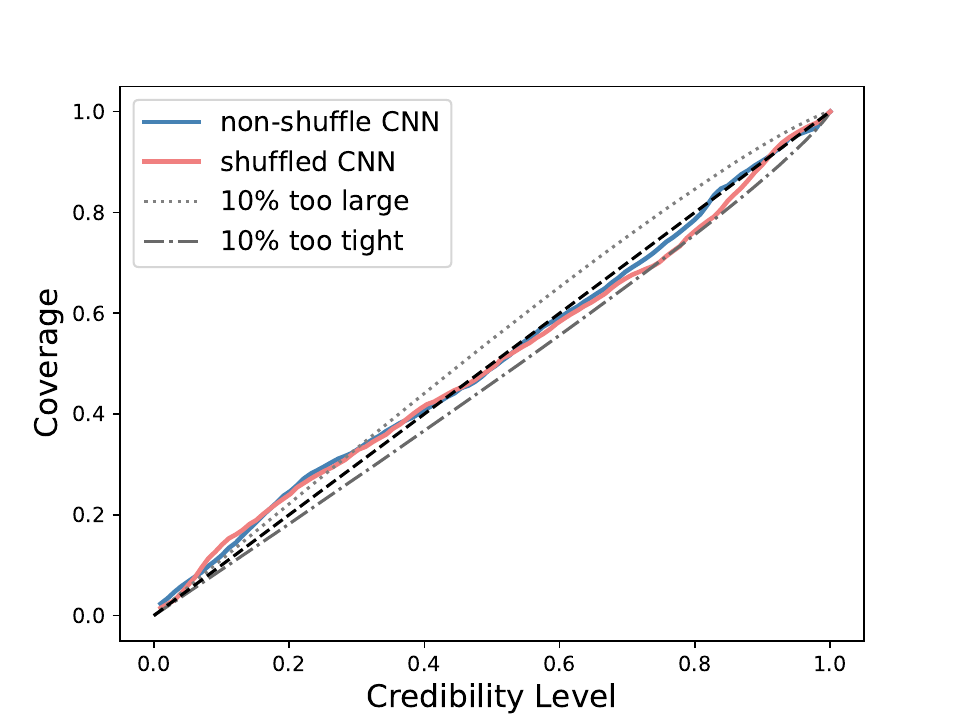}
\caption{Empirical coverage vs credibility level. A well-calibrated posterior closely aligns with the identity line, indicating that the CNN compression is neither overconstrained nor underconstrained. The lines are generated with TARP~\cite{ 2023PMLR..20219256L}, as discussed in Sec.~\ref{sec:sbi}. 
} 
\label{fig:emipircal_coverage} 
\end{figure}
We applied the TARP coverage test with 58 simulations (as there are only 58 independent cosmologies for the $S_8 - \Omega_\mathrm{m}$ plane), and found the posterior is well calibrated as shown in Fig.~\ref{fig:emipircal_coverage}.

A pseudo-code in \textsc{Pytorch} is displayed in Algorithm.~\ref{algorith:1}. Only the lines associated with the random permutation operation are shown. Note that the \textsc{self\_training} variable might not be effective depending on the version of \textsc{Pytorch}. In that case, a customized variable should be used. This is also optional as in some cases random shuffling in both training and validation could boost the overall accuracy. Similar to this, we can also easily define the model to only shuffle the pixels after every N epoch.

\subsection{Experiments to understand the improvement}

\begin{figure}
\centering
\includegraphics[width=\columnwidth]{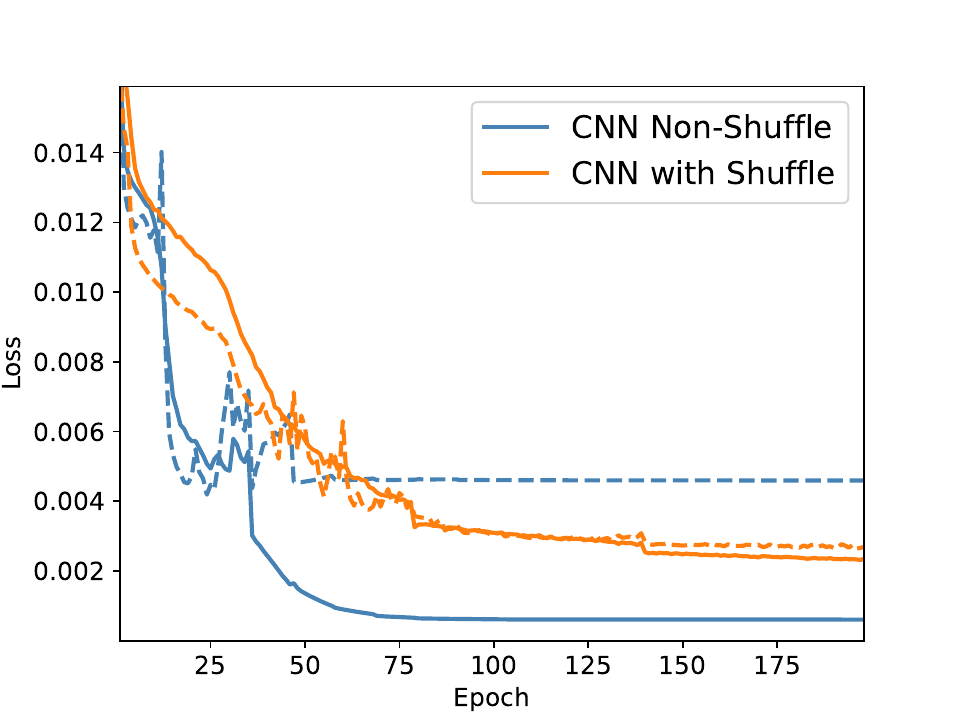}
\caption{The MSE loss vs. training epoch. The solid line shows the training loss while the dashed line shows the validation loss. The blue lines are for the CNN without shuffle and the orange lines are for the shuffled CNN (option 3 in Fig.~\ref{fig:nn_design}). The 6-layer CNN without shuffle can fit the training set very well but the generalization performance is not good as evident from the higher validation loss. This gives an accuracy boost for the models with random shuffling. The loss decay curve shows some differences from case to case, but this is a typical case of how shuffling improves the  CNN. } 
\label{fig:cnn_loss_evolution} 
\end{figure}

To get insights into the source of these improvements, we conducted a series of experiments, varying the models and altering the training datasets. We plot in Fig.~\ref{fig:cnn_loss_evolution} the loss function decay for two models with and without shuffling. The 6-layer CNN non-shuffle model can fit the underlying data very well as suggested by the training loss, but it does not generalize well to the validation or test set. The 4-layer CNN with shuffling, however, can generalize very well, even though it does not fit the training set as well. This demonstrates that the shuffling technique is effective in preventing overfitting. One hypothesis is that the 6-layer non-shuffle model suffers from co-adaptation--a scenario where certain neurons become highly interdependent, leading to overfitting with new test inputs. The shuffling prevents any large-scale outliers from significantly affecting the results.

\begin{figure}
\centering
\includegraphics[width=\columnwidth]{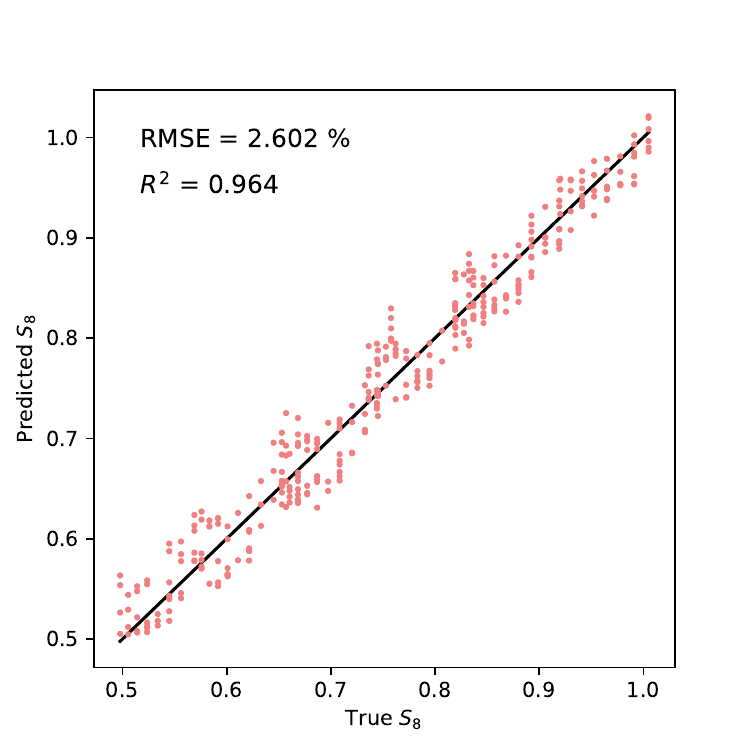}
\caption{Test error for simulation case 1 with the random permutation as the first layers, followed by a $[1 \rightarrow 8]$ convolution layers (8 different kernels to learn). The image is entirely disrupted. The maximum information in each channel is thus the PDF of pixel values. However, if we directly train an MLP using mean, variance, or the full PDF, we get RMSE $\sim$ 7\% at best. This example shows that the convolution operation is a computationally efficient way of summarizing different information across channels.} 
\label{fig:cnn_loss_first_layer_shuffle} 
\end{figure}

One case of special interest is when we directly shuffle the original images. As shown in Fig.~\ref{fig:cnn_loss_first_layer_shuffle}, the result worsens but by less than a factor of 2, which is not as bad as one might expect in the context of standard image analysis, since no structure of the image is retained. However, in cosmology, it is known that the 1-point probability distribution function (PDF) of the image pixels has useful information as it is the PDF of the projected density. In particular, it includes its variance, the smoothed two-point function. The pixel scale in our original image corresponds to a few Mpc at the lens redshift. We test that if we simply measure the average of the entire image, or even use the PDF as summary statistics, the RMSE is $\sim$ 7\%, which is significantly worse. This example shows that even if all the information on scales larger than the single pixel value is destroyed, the convolutional layer (plus subsequent nonlinear operations) can extract more information than just PDFs. We do not pursue this point further in this study. 

Next, we test cases where we removed the large-scale correlation directly from the map. We do this by setting the spherical harmonic coefficients $a_{lm}=0$ for $\ell<\ell_\mathrm{min}$ when making the map for simulation case 2. The results are summarized in Table.~\ref{table:l_cut_tests}. On removing the large-scale correlations, the relative improvement from shuffling is decreased, which makes sense as sample variance in the large-scale modes is also removed by setting $a_{lm}=0$ (so the CNN was not learning the noise in the non-shuffle case). For different shuffling options, the variance is larger but still follows the pattern in Fig.~\ref{fig:cnn_loss_permute_position}.

To check if shuffling in the deeper layers causes the CNN to lose large-scale information, we did the test of adding the lensing power spectrum as a summary statistic to the MLP. We found no improvement in the performance, suggesting that the early layers of the CNN do extract some information on large-scale correlations. We leave for future work whether mode coupling due to nonlinear evolution plays a role in this. 

We note that if the 6-layer CNN model (no shuffling in training) undergoes a post hoc permutation in the final layer during testing, the error only marginally increases from 1.1\% to 1.7\%. This suggests that, typically, the last convolution layer is not optimized to capture spatial correlation but instead functions more as the global average and optimizes performance by its weighting of the different channels. We note that the tests discussed in this section are exploratory; a detailed analysis of regularization methods in cosmology is interesting for future research.

\begin{table}[t]
\renewcommand{\arraystretch}{1.3}
\begin{tabular}{|l|c|c|c|}
\hline
\multicolumn{1}{|l|}{Models}                      & \multicolumn{1}{l|}{RMSE (\%)} & $R^2$  & $\Delta \rm RMSE$\\ \hline
\multicolumn{4}{|c|}{\textbf{Case 2 with NO $\ell_\mathrm{min}$}}                                                   \\ \hline
\multicolumn{1}{|l|}{6-layer CNN (no shuffle)}   & \multicolumn{1}{c|}{3.703}     & 0.934   &   -   \\ \hline
\multicolumn{1}{|l|}{CNN with Shuffle Option 1}   & \multicolumn{1}{c|}{2.946}     & 0.958   &   20.4 \%   \\ \hline
\multicolumn{1}{|l|}{CNN with Shuffle Option 2}   & \multicolumn{1}{c|}{2.914}     & 0.959   &   21.3 \%   \\ \hline
\multicolumn{1}{|l|}{CNN with Shuffle Option 3}   & \multicolumn{1}{c|}{2.579}     & 0.968   &   30.5 \%   \\ \hline

\multicolumn{4}{|c|}{\textbf{Case 2 with $\ell_\mathrm{min} = 76$}}                                      \\ \hline
\multicolumn{1}{|l|}{6-layer CNN (no shuffle)}   & \multicolumn{1}{c|}{4.047}     & 0.923    &  -   \\ \hline
\multicolumn{1}{|l|}{CNN with Shuffle Option 1}   & \multicolumn{1}{c|}{3.837}     & 0.930    & 5.2 \%   \\ \hline
\multicolumn{1}{|l|}{CNN with Shuffle Option 2}   & \multicolumn{1}{c|}{3.261}     & 0.950    & 19.4 \%     \\ \hline
\multicolumn{1}{|l|}{CNN with Shuffle Option 3}   & \multicolumn{1}{c|}{2.953}     & 0.959    & 27.1 \%    \\ \hline

\multicolumn{4}{|c|}{\textbf{Case 2 with $\ell_\mathrm{min} = 148$}}                                    \\ \hline
\multicolumn{1}{|l|}{6-layer CNN (no shuffle)}   & \multicolumn{1}{c|}{4.476}      & 0.903   &  -     \\ \hline
\multicolumn{1}{|l|}{CNN with Shuffle Option 1}   & \multicolumn{1}{c|}{4.095}      & 0.918   &  8.5  \%     \\ \hline
\multicolumn{1}{|l|}{CNN with Shuffle Option 2}   & \multicolumn{1}{c|}{3.549}      & 0.939   &  20.7 \%     \\ \hline
\multicolumn{1}{|l|}{CNN with Shuffle Option 3}   & \multicolumn{1}{c|}{3.240}      & 0.949   &  27.6 \%     \\ \hline

\end{tabular}
\caption{This table summarizes the results for the $S_8$ predictions when we remove the large-scale correlation from the maps by setting the spherical harmonic coefficients $a_{lm}=0$ for $\ell < \ell_\mathrm{min}$. The goal is to understand the shuffle CNN  better as discussed in the text. $\ell_\mathrm{min} = 76$ approximately corresponds to 108 arcmin, and $\ell_\mathrm{min} = 148$ approximately corresponds to 56 arcmin. The column is the percentage improvement over the CNN model without shuffle. In both cases, the overall accuracy is decreased because information is removed from the maps. The relative improvement of shuffling option 1 is also decreased because shuffling at large scales becomes less relevant.}
\label{table:l_cut_tests}
\end{table}

\subsection{Other tests, symmetries, and permutation invariance}

We perform various tests of regularization in comparison with the random permutation layers. In Appendix.~\ref{sec:test_alternatives_to_random_shuffle}, we show 7 additional models that share the same goal as shuffling -- to impose permutation invariance at certain layers. For example, we replaced deep layers with an adaptive average pooling layer that simply takes the average value of the 32x32 features maps to one value. We also tested using other permutation invariant quantities like variance, maximum, and minimum. The results suggest that they are not as effective as the shuffling option. In Appendix.~\ref{sec:other_regularization} we compare with two other regularization schemes \textsc{Dropout} and batch normalization. We showed that they negatively affect the accuracy in our case. As a supplementary test, we show in Appendix.~\ref{sec:additional_test} the results for different noise levels and training set sizes. Across various test scenarios, the addition of random permutations consistently elevates the generalization accuracy. 

Constraining neural networks with physical symmetries has been shown to help the training and the accuracy of the model. In cosmology particularly, translation and rotation symmetries are key ingredients to improve performance. Previous work that imposes these symmetries in normalizing flows~\cite{Dai:2022dso} and graph neural network~\cite{2022OJAp....5E..18M} demonstrate great potential for both generative models and models used for parameter inference. The scattering wavelet transform~\cite{https://doi.org/10.1002/cpa.21413} is powerful for cosmological fields partially due to its translation invariance construction. The convolution operation is translation equivariant by definition. However, note that the max-pooling operation breaks the translation equivariance so the CNN is not rigorously translation invariant. Investing in architectures suitable for cosmology with translational~\cite{2019arXiv190411486Z} and rotational symmetry~\cite{2016arXiv160207576C, 2019arXiv191108251W} is a promising future direction -- we leave such investigations for future work. 

Permutation invariance, although less relevant in CNN, is very common for Graph Neural Networks (GNN)~\cite{2019arXiv190100596W}. \textsc{Janossy Pooling}~\cite{2018arXiv181101900M} employs a random sample of a subset of possible permutations as an alternative pooling operation. Such tweaks to the training process can also be found in CNNs. In particular, in \textsc{ShuffleNet}~\cite{2017arXiv170701083Z} a channel shuffle preceding a group convolution is employed. A more extreme case can be found in~\cite{2016arXiv160500055X}, where the authors disturb the loss layer by directly giving wrong labels, and they show it boosts the neural network's ability to learn more general features.

In this work, we shuffle the feature map for every epoch but only for training. Other variations of random permutation are also worth exploring -- for example, we tested shuffle every 10 epochs, which allows the convolution layer to learn the same map within these 10 epochs. The test error is similar to that of shuffling every epoch. One can also introduce a probability distribution where say only 10\% of the time the feature map is shuffled. We expect that shuffling every $N$ epoch to be a more general solution that can be optimized for other cases.


\section{Discussions and Conclusions}
\label{sec:Conclusion}

CNNs are the most extensively developed deep learning approach for image analysis. In recent years they have also been applied for inferring cosmological parameters from weak lensing and other cosmological maps. We show that for the characteristics of cosmological fields like lensing mass maps, variations of the standard CNN architecture can lead to improved performance.  We have presented a set of regularization and data augmentation methods and quantified the performance improvement for simulated lensing surveys. The simulated surveys mimic the Dark Energy Survey parameters in both statistical and systematic uncertainties, but we also consider the lower shape noise levels corresponding to Stage-IV surveys. The explorations in this paper are not definitive as the detailed implementation should be adapted to the specific applications. 

We use random permutations (shuffling operation) in the deep layers of CNN as a novel regularization and data augmentation technique. We also use a mix of maximum and average pooling for varying layer depths. These simple modifications can enhance the performance compared to traditional CNN designs that have been adapted from image classification. Figure.~\ref{fig:posterior} demonstrates the improvement in parameter inference. We also show in Fig.~\ref{fig:cnn_loss_evolution} that it can be expected to have better generalization accuracy. While the design requires optimization for specific problems, incorporating a random permutation layer emerges as a promising approach for cosmological fields. An open question that we have addressed only partially is whether CNNs like the ones we have tested lose some large-scale information. It is possible that the early layers of the CNN, prior to shuffling, have extracted the available signal. We find some evidence for this when considering the power spectrum -- adding the power spectrum, including the large scales (small angular wavenumbers), to the MLP does not improve the performance further. We leave a detailed investigation for future work. 

With the upcoming Stage-IV surveys (Euclid, Rubin-LSST, and Roman), we expect a substantial improvement in the quality and size of lensing mass maps. Deep learning approaches along with higher-order statistics and field-level inference can extract the vast amount of information from these maps that goes beyond standard 2-point statistics. While CNNs may not capture all the information contained in cosmological maps, they offer some clear advantages. As a model-specific data compressor, CNNs are flexible for a lot of different settings. For example, if the data has multiple tomographic bins as in current weak lensing surveys, CNNs can take them as different channels and capture the cross-channel information. If one prefers to analyze data with a shear field instead of the convergence field, one can also treat the two components as different channels and avoid the complexity of reconstructing the convergence map. Hence exploring improved performance with CNNs is likely to be of value in cosmological applications. 

\textbf{Caveats and Future work}: This paper has focused on testing a set of regularization and data augmentation schemes with a simple CNN design. Given that this approach is inspired by our understanding that cosmological fields are noisy on scales approaching the survey size, we anticipate that the random permutation layer, when applied at appropriate scales, could be synergized with more complex architectures. For instance, similar shuffling operations can be easily tested with Vision Transformer~\cite{Hwang:2023oob} or Graph-based Neural Networks~\cite{2019A&C....27..130P}.

It is also interesting to test if this regularization and data augmentation method can be effective in other cosmological fields such as the overdensity field~\cite{2020arXiv201105992V, Villaescusa-Navarro:2021pkb}, 21-cm maps~\cite{2023MNRAS.tmp.2817N}, and secondary anisotropies in CMB temperature or polarization maps~\cite{Casas:2022teu, Munchmeyer:2019kng}.

A useful direction for future research is the optimization of the shuffling operation. One possibility is to introduce a hyperparameter that modulates the probability of performing the shuffle, which is a technique used in~\cite{Swiderski2022}. Our limited explorations are conservative in the sense that we did not tune hyperparameters, so there is room for improvement by tuning all the hyperparameters for different variations of the noise level. Additionally, exploring other randomness-incorporating structures, such as random shifting~\cite{ijcai2017p486} or randomly wired CNNs~\cite{2019arXiv190401569X}, could also offer valuable insights. 

Another important direction for future research is in the interpretability of CNNs when applied to cosmological data. Despite their proven efficacy in regression problems, a persistent challenge with CNNs is the 'black box' nature of their decision-making processes. Understanding how these networks derive their conclusions is important, not just for validating and improving accuracy, but also for gaining deeper insights into the underlying physics. Previous studies~\cite{2020PhRvD.102l3506M, Ntampaka:2021scs} show intriguing results by using saliency maps~\cite{2013arXiv1312.6034S, 2014arXiv1412.6806S} (see also a recent study~\cite{2023arXiv231016316Y} using a `sum-of-parts' approach to interpretability). Architectures designed to relate the neural network results to physical qualities like the \textit{N}-point correlation function are also promising~\cite{Miles2021, Gong:2024jsw}. It will be interesting to see how the results change with the presence of random permutation layers.
\section*{Acknowledgements}

We thank Shubh Agrawal, Pratik Chaudhari, Cyrille Doux, Rafael Gomes, Mike Jarvis, Amrut Nadgir, Shivam Pandey, Helen Qu, Dimitrios Tanoglidis, and Eric Wong for useful discussions. We are especially grateful to Tomek Kacprzak and his collaborators for generating and making available the \textsc{DarkGrid} simulations used in this work. This research used resources of the National Energy Research Scientific Computing Center (NERSC), a U.S. Department of Energy Office of Science User Facility located at Lawrence Berkeley National Laboratory, operated under Contract No. DE-AC02-05CH11231. B.J. and M.G. are supported in part by the US Department of Energy grant DE-SC0007901.

\bibliography{ref_short.bib}

\appendix

\section{Test pooling choices}\label{sec:test_pooling}
\begin{table}[t]
\renewcommand{\arraystretch}{1.3}
\begin{tabular}{|l|c|c|}
\hline
Models                 & \multicolumn{1}{l|}{RMSE (\%)} & $R^2$\\ \hline
\multicolumn{3}{|c|}{\textbf{simulation case 1}}                         \\ \hline
All Max Pooling   & 2.174                     & 0.975           \\ \hline
All Avg Pooling   & 1.308                     & 0.991           \\ \hline
\textbf{Mixed Pooling}    & \textbf{1.187}                     & 0.992           \\ \hline
\multicolumn{3}{|c|}{\textbf{simulation case 2}}                         \\ \hline
All Max Pooling   & 4.764                    & 0.890           \\ 
\hline
\textbf{All Avg Pooling}   & \textbf{3.61 }                    & 0.937              \\ \hline
Mixed Pooling    & 3.703                     & 0.934         \\ \hline

\multicolumn{3}{|c|}{\textbf{simulation case 2 with $2 n_{\mathrm{eff}}$}}                         \\ \hline
All Max Pooling   & 4.5820                     & 0.898           \\ 
\hline
\textbf{All Avg Pooling}   & \textbf{2.520}                       & 0.969              \\ \hline
Mixed Pooling    & 3.079                     & 0.954           \\ \hline
Mixed Pooling-v2    & 2.699                     & 0.964           \\ \hline

\multicolumn{3}{|c|}{\textbf{simulation case 2 with $4 n_{\mathrm{eff}}$}}                         \\ \hline
All Max Pooling   & 3.526                     & 0.939           \\ 
\hline
All Avg Pooling   & 2.042                       & 0.980              \\ \hline
Mixed Pooling    & 2.380                   & 0.972           \\ \hline
\textbf{Mixed Pooling-v2}    & \textbf{1.683}                   & 0.986           \\ \hline

\hline

\end{tabular}
\caption{Summary of comparing pooling choices in different simulation cases when applied to $S_8$. Mixed Pooling stands for the 4max+2avg as shown in Fig.~\ref{fig:nn_design}. Mixed Pooling-v2 denotes 2max+4avg. Note that for simulation case 2 the all-avg is slightly better than mix-pooling, but the all-avg does not work at all for inferring $\Omega_\mathrm{m}$. Hence, we use mixed pooling (4max+2avg) as the baseline choice for this work.}

\label{table:test_hybrid_pooling}
\end{table}

The experiments involving various pooling options are detailed in Table \ref{table:test_hybrid_pooling}. The outcomes vary between simulation case 1 and 2 and with the level of shape noise but not in a straightforward way. This quantitive comparison is restricted to $S_8$. We selected mixed pooling as the baseline since the all-average approach is consistently ineffective for training on $\Omega_\mathrm{m}$. Although this investigation is far from comprehensive due to computational resources limitations, it shows that different choices in pooling can impact the outcomes.

\section{Testing Alternatives to Random Permutation Layer}
\label{sec:test_alternatives_to_random_shuffle}
\begin{table}[t]
\renewcommand{\arraystretch}{1.3}
\begin{tabular}{|l|c|c|}
\hline
Models                 & \multicolumn{1}{l|}{RMSE (\%)} & $R^2$\\ \hline
Best Model with Shuffle     & 0.865                     & 0.996           \\ \hline
Test Model 1   & 1.235                     & 0.992                 \\ \hline
Test Model 2   & 1.126                     & 0.993                \\ \hline
Test Model 3   & 1.147                     & 0.993                \\ \hline
Test Model 4   & 1.263                     & 0.991                \\ \hline
Test Model 5   & 1.419                     & 0.989                \\ \hline
Test Model 6   & 2.316                     & 0.971                 \\ \hline
Test Model 7   & 1.84                      & 0.982                 \\ \hline
\end{tabular}
\caption{Different models that we test as alternatives to random permutation layers, in comparison with the best model with shuffle (option 2). The description for each model is listed in Appendix.~\ref{sec:test_alternatives_to_random_shuffle}. None of these models performs better than even the 6-layer CNN model without shuffling. This test shows that the effect of random permutation layers cannot be replaced by such permutation invariant statistics. }
\label{table:test_models}
\end{table}

Since introducing random permutation layers improves performance, we consider other ways to impose permutation invariance. A large average pooling over the entire feature map, for example, is permutation invariant. One can also add other permutation-invariant summary statistics such as the variance, maximum, or minimum. This is more common in Graph Neural Networks, where permutation invariance is crucial due to the data structure. For example, in~\cite{2023ApJ...955..131W}, the input of the MLP layers is the collection of sum, mean, max, and min. 

Here, we test the following possibilities:
\begin{description}
  \item[$\bullet$ Test Model 1-3. Large Avg] These test models replace the random permutation layer and the subsequent convolution layer with a large average pooling. Test model 1-3 corresponds to models that have 5, 4, and 3 convolution layers. Note that the following MLP also changes in size because the channel number changes.
  \item[$\bullet$ Test Model 4. Mean + Variance] This model employs a large average pooling layer and additionally calculates the variance of the 8x8 pixel blocks for each channel. These are then concatenated as the input for the MLP.
  \item[$\bullet$ Test Model 5. Mean + Sum + Max + Min] This model is an extension of Test Model 4, using the Mean (Avg), Sum, Maximum, and Minimum of the current channel as the inputs of the MLP.
  \item[$\bullet$ Test Model 6. Large Flat Layer] In this model, the final average pooling is omitted. Instead, a flattening operation is used to compile all the information. The input size of MLP is increased to $128\times8\times8=8192$. We also tested other channel numbers with flattening in the end and found similar results.
  \item[$\bullet$ Test Model 7. Patchify into smaller CNNs] For this model, we patchify the original image to 8x8 smaller images of size 64x64. We forward these smaller patches to the same weight-sharing convolutional layers and then take the average as the input of the MLP layers.
\end{description}
Note that all of the models above do not involve random permutation. Models 1 to 5 are strictly permutation invariant at corresponding layers. The test errors are summarized in Table~\ref{table:test_models}, and they do not help the 6-layer CNN model as the random permutation layers. This is a model-specific result. Following the idea of random shuffling at large scales, we anticipate some of the choices above could increase accuracy in certain simulation settings. However, the random permutation layer remains a possibility when designing neural networks for cosmological fields.

For Model 7, we patchify the original image into smaller patches, and we tried various combinations of patch sizes, convolutional layers, and MLP sizes. This choice shares the same idea of making the corresponding scales permutation invariant. However, the modes between each patch are not captured by the convolution operations, and thus much less information is available. The RMSE is 1.84, significantly larger than the 4-layer shuffled model or the 6-layer non-shuffle model. Note that this experiment is conducted with simulation case 1, and we expect this method to work worse for simulation case 2 in the presence of masks.

\section{Comparison with Standard Regularization}\label{sec:other_regularization}
\subsection{\textsc{Dropout}}\label{sec:Dropout}

\textsc{Dropout} is a widely recognized regularization technique for mitigating overfitting in machine learning. Essentially, \textsc{Dropout} combats co-adaptation by randomly deactivating a portion of the neurons. Despite its effectiveness, \textsc{Dropout} is not universally applicable, especially in more recent and complex computer vision models like ResNet~\cite{2015arXiv151203385H}.

We show in Table~\ref{table:regularization} three different use of \textsc{Dropout}:
\begin{description}
  \item[\textsc{\textsc{Dropout}}-v1] One \textsc{Dropout} layer before the MLP of the 6-layer CNN model. This test is to see if masking some of the embedding representations can prevent overfitting of MLP
  \item[\textsc{Dropout}-v2] Three \textsc{Dropout} layers after each linear+activation function in MLP, namely fc 7-9 in Fig.~\ref{fig:nn_design}. This is the more traditional use of \textsc{Dropout}, where all the representations from the convolution layer are kept, but the use \textsc{Dropout} directly prevents the overfitting of the MLP regression.
  \item[\textsc{Dropout}-v3] A 2D \textsc{Dropout} layer before convolution 6, where the random permutation layer is applied. This test is to make a direct comparison to the permutation layer, but instead of shuffling the positions, some patches are masked out during training.
\end{description}
For all three tests, we use a \textsc{Dropout} rate of 0.2. We found no improvement over the 6-layer CNN model.

\begin{table}[t]
\renewcommand{\arraystretch}{1.3}
\begin{tabular}{|lcc|}
\hline
\multicolumn{1}{|l|}{Models}                     & \multicolumn{1}{l|}{RMSE (\%)} & $R^2$ \\ \hline
\multicolumn{1}{|l|}{Best Model with Shuffle}         & \multicolumn{1}{c|}{0.865    }     & 0.996                \\ \hline
\multicolumn{3}{|c|}{\textbf{\textsc{Dropout}}}                                                                            \\ \hline
\multicolumn{1}{|l|}{All Avg w/ \textsc{Dropout}-v1}         & \multicolumn{1}{c|}{1.214}     & 0.992                \\ \hline
\multicolumn{1}{|l|}{Mixed Avg/Max w/ \textsc{Dropout}-v1}   & \multicolumn{1}{c|}{1.115}     & 0.993                \\ \hline
\multicolumn{1}{|l|}{Mixed Avg/Max w/ \textsc{Dropout}-v2}   & \multicolumn{1}{c|}{1.563}     & 0.987                \\ \hline
\multicolumn{1}{|l|}{Mixed Avg/Max w/ \textsc{Dropout}-v3}   & \multicolumn{1}{c|}{1.165}     & 0.993                \\ \hline
\multicolumn{3}{|c|}{\textbf{\textsc{BatchNorm} before activation}}                                                         \\ \hline
\multicolumn{1}{|l|}{All Avg w/ BN}       & \multicolumn{1}{c|}{1.664}     & 0.985                \\ \hline
\multicolumn{1}{|l|}{Mixed Avg/Max w/ BN} & \multicolumn{1}{c|}{2.780}      & 0.959                 \\ \hline
\multicolumn{3}{|c|}{\textbf{\textsc{BatchNorm} after activation}}                                             \\ \hline
\multicolumn{1}{|l|}{All Avg w/ BN}       & \multicolumn{1}{c|}{2.169}        & 0.975               \\ \hline
\multicolumn{1}{|l|}{Mixed Avg/Max w/ BN}  & \multicolumn{1}{c|}{1.890}        & 0.981                 \\ \hline
\end{tabular}
\caption{Test models with two other regularization methods \textsc{Dropout} and \textsc{BatchNorm} when applied to simulation case 1 and $S_8$. We also tested  applying \textsc{BatchNorm} before and after each activation function. See Appendix.~\ref{sec:other_regularization} for the details of where these regularization layers are used. As in Table~\ref{table:test_models}, none of these methods do as well as the shuffle CNN. 
}
\label{table:regularization}
\end{table}

\subsection{\textsc{BatchNorm}}\label{sec:Batch_norm}
\begin{table}[t]
\centering
\begin{tabular}{cccc}
\hline
Layer           & Kernel size & Stride & Output dimensions       \\ \hline
(Input)         &             &        & (4 $\times$ 512 $\times$ 512) \\
Convolution     & 5 $\times$ 5 & 2      & $(n_{ch}/2) \times 254 \times 254$ \\
Convolution     & 5 $\times$ 5 & 2      & $n_{ch} \times 125 \times 125$ \\
Residual block  &             &        &                          \\
\vdots          & \vdots      & \vdots & \vdots                  \\
Residual block  &             &        &                          \\
Pooling         & 2 $\times$ 2 & 2      & $n_{ch} \times 62 \times 62$ \\
Convolution     & 3 $\times$ 3 & 1      & $(2n_{ch}) \times 60 \times 60$ \\
Pooling         & 2 $\times$ 2 & 2      & $(2n_{ch}) \times 30 \times 30$ \\
Convolution     & 3 $\times$ 3 & 1      & $(4n_{ch}) \times 28 \times 28$ \\
Pooling         & 2 $\times$ 2 & 2      & $(4n_{ch}) \times 14 \times 14$ \\
Convolution     & 3 $\times$ 3 & 1      & $(8n_{ch}) \times 12 \times 12$ \\
Pooling         & 2 $\times$ 2 & 2      & $(8n_{ch}) \times 6 \times 6$ \\
Convolution     & 3 $\times$ 3 & 1      & $(16n_{ch}) \times 4 \times 4$ \\
Pooling         & 4 $\times$ 4 & —      & $(16n_{ch}) \times 1 \times 1$ \\
Linear          & —           & —      & 256                        \\
ReLU            & —           & —      & 256                        \\
Linear          & —           & —      & 1                          \\ \hline
\end{tabular}
\caption{The architecture used in~\cite{2022MNRAS.511.1518L}, except for the difference in the input and output size. The RMSE for simulation case 2 is 3.776\%, which is similar to our  non-shuffle CNN model as shown in Fig.~\ref{fig:cnn_loss_case_2}.}
\label{table:resnet}
\end{table}

Batch Normalization (\textsc{BatchNorm})~\cite{2015arXiv150203167I} is another very powerful regularization scheme that is commonly used in Machine Learning and has been adopted in previous weak lensing analyses such as~\cite{2023MNRAS.521.2050L, Ribli2019, 2022MNRAS.511.1518L,  2023MNRAS.521.2050L}, but not in~\cite{Fluri2018}. \textsc{BatchNorm} essentially whitens the output activation function and thus alleviates the internal covariance shift problem~\footnote{See \hyperlink{https://joelouismarino.github.io/posts/2017/08/statistical_whitening/}{https://joelouismarino.github.io/posts/2017/08/statistical-whitening/} for a detailed explanation on statistical whitening and data normalization}, which is common in most of the neural network design.

\begin{table}[t]
\renewcommand{\arraystretch}{1.3}
\begin{tabular}{|lcc|}
\hline
\multicolumn{1}{|l|}{Models}               & \multicolumn{1}{l|}{RMSE (\%)} & $R^2$ \\ \hline
\multicolumn{3}{|c|}{\textbf{simulation case 1 with $2 n_\mathrm{eff}$}}                                      \\ \hline
\multicolumn{1}{|l|}{non-shuffle 6-layer CNN}         & \multicolumn{1}{c|}{0.974}            & 0.996         \\ \hline
\multicolumn{1}{|l|}{CNN with Shuffle Option 1}   & \multicolumn{1}{c|}{0.740}     & 0.998         \\ \hline
\multicolumn{1}{|l|}{CNN with Shuffle Option 2}   & \multicolumn{1}{c|}{\textbf{0.684}}     & 0.998         \\ \hline

\multicolumn{3}{|c|}{\textbf{simulation case 1 with $0.5 n_\mathrm{eff}$}}                                    \\ \hline
\multicolumn{1}{|l|}{non-shuffle 6-layer CNN}       & \multicolumn{1}{c|}{1.537}             & 0.988          \\ \hline
\multicolumn{1}{|l|}{CNN with Shuffle Option 1} & \multicolumn{1}{c|}{\textbf{1.275}}      & 0.991          \\ \hline
\multicolumn{1}{|l|}{CNN with Shuffle Option 2} & \multicolumn{1}{c|}{1.833}      & 0.982          \\ \hline

\multicolumn{3}{|c|}{\textbf{simulation case 1 with no shape noise}}                                          \\ \hline
\multicolumn{1}{|l|}{non-shuffle 6-layer CNN}       & \multicolumn{1}{c|}{0.561}             & 0.998          \\ \hline
\multicolumn{1}{|l|}{CNN with Shuffle Option 1} & \multicolumn{1}{c|}{\textbf{0.424}}      & 0.999          \\ \hline
\multicolumn{1}{|l|}{CNN with Shuffle Option 2} & \multicolumn{1}{c|}{0.565}      & 0.998          \\ \hline

\multicolumn{3}{|c|}{\textbf{simulation case 2 with half training data}}                                      \\ \hline
\multicolumn{1}{|l|}{non-shuffle 6-layer CNN}       & \multicolumn{1}{c|}{4.097}        & 0.912                   \\ \hline
\multicolumn{1}{|l|}{CNN with Shuffle Option 1}  & \multicolumn{1}{c|}{3.594}        & 0.932           \\ \hline
\multicolumn{1}{|l|}{CNN with Shuffle Option 2}  & \multicolumn{1}{c|}{3.478}        & 0.936           \\ \hline
\multicolumn{1}{|l|}{CNN with Shuffle Option 3}  & \multicolumn{1}{c|}{\textbf{2.921}}        & 0.955           \\ \hline

\multicolumn{3}{|c|}{\textbf{simulation case 2 with $2 n_\mathrm{eff}$}}                                      \\ \hline
\multicolumn{1}{|l|}{non-shuffle 6-layer CNN (mixed-pooling)}       & \multicolumn{1}{c|}{3.079}        & 0.954                   \\ \hline
\multicolumn{1}{|l|}{non-shuffle 6-layer CNN (avg-pooling)}       & \multicolumn{1}{c|}{2.520}        & 0.969                   \\ \hline
\multicolumn{1}{|l|}{CNN with Shuffle Option 1}  & \multicolumn{1}{c|}{2.520}        & 0.969           \\ \hline
\multicolumn{1}{|l|}{CNN with Shuffle Option 2}  & \multicolumn{1}{c|}{2.179}        & 0.977           \\ \hline
\multicolumn{1}{|l|}{CNN with Shuffle Option 3}  & \multicolumn{1}{c|}{\textbf{2.013}}        & 0.980           \\ \hline

\multicolumn{3}{|c|}{\textbf{simulation case 2 with $4 n_\mathrm{eff}$}}                                      \\ \hline
\multicolumn{1}{|l|}{non-shuffle 6-layer CNN (mixed-pooling-v2)}       & \multicolumn{1}{c|}{1.683}        & 0.9886                   \\ \hline
\multicolumn{1}{|l|}{non-shuffle 6-layer CNN (avg-pooling)}       & \multicolumn{1}{c|}{2.042}        & 0.980                   \\ \hline
\multicolumn{1}{|l|}{CNN with Shuffle Option 1}  & \multicolumn{1}{c|}{1.994}        & 0.981           \\ \hline
\multicolumn{1}{|l|}{CNN with Shuffle Option 2}  & \multicolumn{1}{c|}{1.748}        & 0.985           \\ \hline
\multicolumn{1}{|l|}{CNN with Shuffle Option 3}  & \multicolumn{1}{c|}{\textbf{1.410}}        & 0.990           \\ \hline

\end{tabular}
\caption{Additional tests varying noise level and size of the training set. The reported RMSE is for $S_8$. The shuffling operation consistently improves the accuracy when comparing non-shuffle 6-layer CNN and shuffle option 1, which only differs by one random permutation layer. Note that the best position of the shuffling operation changes from the fiducial test of the paper. }
\label{table:additionl_test}
\end{table}

However, \textsc{BatchNorm} is not a once and for all solution. As a simple example, putting \textsc{BatchNorm} before or after the activation function is non-trivial, with either choice differently affects the goal of whitening~\footnote{See \hyperlink{http://torch.ch/blog/2016/02/04/resnets.html}{http://torch.ch/blog/2016/02/04/resnets.html} for a discussion on different effects of \textsc{BatchNorm} before or after the activation. }. The issue is further complicated with regularization schemes such as \textsc{Dropout}~\cite{2018arXiv180105134L}. In fact, a large number of architectures find that when combining the two most powerful regularization schemes \textsc{Dropout} and \textsc{BatchNorm}, the accuracy often decreases. 

In this work, we did not adopt any \textsc{BatchNorm} in our architecture, and we show in Table~\ref{table:regularization} that it negatively affects the results. Note that we use the default setting that enables affine parameters, which are the only two learnable parameters. However, this complicates the training in the mini-batch setting because they have a large effect on the whitened output. 

Again, this choice is only specific to optimizing our problem, which can be different if the data and other assumptions change. For instance, if one uses data augmentation with less Gaussian noise injection level, or studies the case with all cosmological parameters varying, the internal large variance of the input data can make \textsc{BatchNorm} necessary. We do, nevertheless, expect the permutation layer to improve the generalization performance in the presence of \textsc{BatchNorm}. 

\section{Comparison with ResNet used in previous work}\label{sec:comsparison_with_resnet}

We also test an architecture with a residual connection. The architecture is summarized in Table~\ref{table:resnet}, which is the same as in~\cite{2022MNRAS.511.1518L,  2023MNRAS.521.2050L} except the input channel being 4 for our 4 tomography bin and output number being 1 since we are training one parameter per network. The residual block consists of two convolution layers and a skip connection, similarly in~\cite{2015arXiv151203385H}. We use 10 residual blocks and $n_\mathrm{ch}=10$ as suggested in~\cite{2022MNRAS.511.1518L}. We find no difference between the ResNet and our 6-layer CNN model without shuffle. For simplicity, we did not further optimize this design or test with random permutation layers with this structure. This test suggests that the residual connection might not be necessary for our simulations. We stress again that the comparison is not exact because our simulations are very different. Even though the input sizes are both 512x512, the resolution is very different, with those in~\cite{2015arXiv151203385H} being 0.87 arcmin. However, this result suggests that the proposed random permutation layer is optimizing the neural network for cosmological fields in a different way.

\section{Additional Tests with Different Noise Level and Size of Training Set}\label{sec:additional_test}
We provide some additional tests varying the noise level and training data size. As summarized in Table~\ref{table:additionl_test}, the shuffling operation consistently boosts the generalization accuracy of the model. Note that the 6-layer CNN should be compared with CNN with shuffle option 1 as they have the same structure and trainable parameters, but only differ by one random permutation layer. In these tests, the other option is not always better than option 1, but CNN with Shuffle Option 1 is always better than the 6-layer CNN model.

\end{document}